\documentclass[pra,twocolumn,aps,superscriptaddress,nofootinbib]{revtex4-1}
\usepackage{amsfonts}
\usepackage{mathtools}
\usepackage[svgnames,table]{xcolor}
\usepackage[T1]{fontenc}
\usepackage{txfonts}
\usepackage{graphicx}
\usepackage{lineno}
\usepackage{gensymb}
\usepackage{subcaption}

\usepackage{hyperref}
\hypersetup{
        colorlinks=true,
        citecolor=SteelBlue,
        filecolor=LimeGreen,
        linkcolor=SlateBlue,
        urlcolor=MediumPurple
}

\usepackage{dcolumn}
\usepackage{bm}
\usepackage{amsmath}
\usepackage{breqn}
\usepackage[bbgreekl]{mathbbol}
\usepackage{bbm}
\usepackage{mathrsfs}
\usepackage[permil]{overpic}

\makeatletter
\let\cat@comma@active\@empty
\makeatother
\begin{document}
\preprint{APS/123-QED}
\title{Active platform stabilisation with a 6D seismometer}
\author{Amit Singh Ubhi}
\affiliation{Institute for Gravitational Wave Astronomy, School of Physics and Astronomy, University of Birmingham, Birmingham B15 2TT, United Kingdom}
\author{Leonid Prokhorov}
\affiliation{Institute for Gravitational Wave Astronomy, School of Physics and Astronomy, University of Birmingham, Birmingham B15 2TT, United Kingdom}
\author{Sam Cooper}
\affiliation{Institute for Gravitational Wave Astronomy, School of Physics and Astronomy, University of Birmingham, Birmingham B15 2TT, United Kingdom}
\author{Chiara Di Fronzo}
\affiliation{Institute for Gravitational Wave Astronomy, School of Physics and Astronomy, University of Birmingham, Birmingham B15 2TT, United Kingdom}
\author{John Bryant}
\affiliation{Institute for Gravitational Wave Astronomy, School of Physics and Astronomy, University of Birmingham, Birmingham B15 2TT, United Kingdom}
\author{David Hoyland}
\affiliation{Institute for Gravitational Wave Astronomy, School of Physics and Astronomy, University of Birmingham, Birmingham B15 2TT, United Kingdom}

\author{Alexandra Mitchell}
\affiliation{Department of Physics and Astronomy, VU Amsterdam, 1081 HV, Amsterdam, The Netherlands}
\affiliation{Dutch National Institute for Subatomic Physics, Nikhef, 1098 XG, Amsterdam, Netherlands}

\author{Jesse van Dongen}
\affiliation{Department of Physics and Astronomy, VU Amsterdam, 1081 HV, Amsterdam, The Netherlands}
\affiliation{Dutch National Institute for Subatomic Physics, Nikhef, 1098 XG, Amsterdam, Netherlands}

\author{Conor Mow-Lowry}
\affiliation{Institute for Gravitational Wave Astronomy, School of Physics and Astronomy, University of Birmingham, Birmingham B15 2TT, United Kingdom}
\affiliation{Department of Physics and Astronomy, VU Amsterdam, 1081 HV, Amsterdam, The Netherlands}
\affiliation{Dutch National Institute for Subatomic Physics, Nikhef, 1098 XG, Amsterdam, Netherlands}

\author{Alan Cumming}
\affiliation{Institute for Gravitational Wave Research, School of Physics and Astronomy, University of Glasgow, Glasgow G12 8QQ, United Kingdom}
\author{Giles Hammond}
\affiliation{Institute for Gravitational Wave Research, School of Physics and Astronomy, University of Glasgow, Glasgow G12 8QQ, United Kingdom}
\author{Denis Martynov}
\affiliation{Institute for Gravitational Wave Astronomy, School of Physics and Astronomy, University of Birmingham, Birmingham B15 2TT, United Kingdom}

\date{\today}

\begin{abstract}

We demonstrate the control scheme of an active platform with a six degree of freedom (6D) seismometer. The inertial sensor simultaneously measures translational and tilt degrees of freedom of the platform and does not require any additional sensors for the stabilisation. We show that a feedforward cancellation scheme can efficiently decouple tilt-to-horizontal coupling of the seismometer in the digital control scheme. We stabilise the platform in the frequency band from 250\,mHz up to 10\,Hz in the horizontal degrees of freedom and achieve a suppression factor of 100 around 1\,Hz. Further suppression of ground vibrations was limited by the non-linear response of the piezo actuators of the platform and by its limited range ($5\,\mu$m). In this paper we discuss the 6D seismometer, its control scheme, and the limitations of the test bed.


\end{abstract}

\maketitle

\section{Introduction}

The LIGO~\cite{Aasi2015} and Virgo~\cite{Acernese2014} detectors have made a number of gravitational wave detections from massive compact objects~\cite{Abbott2016, Abbott2019GWTC1, Abbott2021GWTC2, Nitz2021}. Sources of these waves range from two recent neutron star black hole systems~\cite{Abbott2021}, and binary black holes~\cite{Collaboration2016, Collaboration2017,Abbott2020a}, with one detection of an intermediate mass black hole of mass $\sim 150\,\rm M_{\odot}$~\cite{Abbott2020a}. A multimessenger event was also observed from a binary neutron star merger which verified localisation and decreased the false alarm rate of the detection~\cite{Collaboration2017a}.

Low frequency sensitivity of the detectors determine the likelihood of observing more massive systems such as intermediate mass black hole binaries between $100-1000M_\odot$ aswell as providing early warning signals. The merger time of binary systems scale with frequency as $f^{-8/3}$, enabling opportunities for multimessenger detections. For the LIGO detectors, these signals are cloaked by the non-stationary control noise of the isolation scheme of the core optics~\cite{Yu2018,Buikema2020,Martynov_Noise_2016}. The LIGO isolation scheme consists of a four stage pendulum suspended from state of the art two stage twelve axis platforms for the detectors' core optics~\cite{Matichard_2015,MATICHARD2015287,MATICHARD2015273}. Despite the orders of magnitude suppression achieved, the angular controls for the core optics limit the detectors' sensitivity below 30\,Hz~\cite{Dooley_ASC_2013,Barsotti_ASC_2010}. 

Improved sensing of the isolated platforms would reduce the input motion to the suspension chain, reducing the injection of noise from the local damping on the optics. Suppression of platform tilt is limited by the lack of absolute rotation sensors on the platforms. The platform tilt also plagues the translational readout with an unfavourable coupling of $g/\omega^2$~\cite{Matichard2015, Matichard2016},where $g$ is the local gravitational acceleration and $\omega$ the angular frequency. Investigations into improved sensing of the platforms are being explored by a number of groups who develop novel inertial sensors. Krishna Venkateswara at the University of Washington has employed the out of vacuum beam rotation sensor (BRS)~\cite{Venkateswara2014} at LIGO for feedforward correction of translational sensors. The University of Washington is also developing an in vacuum cylindrical rotation sensor (CRS). The University of Western Australia have developed the ALFRA rotational accelerometer which has the advantage of multi-orientation such that it can also be mounted vertically~\cite{McCann_2021}. Optical gyroscopes have also been investigated at Caltech and MIT which make use of the Sagnac effect to measure absolute rotation~\cite{Korth_Gyro_15,Martynov_Gyro_2019}. Further improvements to low noise translational inertial sensing have been demonstrated by the Nikhef and VU groups in Amsterdam~\cite{Heijningen_2018}, and the Belgium China collaboration~\cite{Collette2015,Ding2022} with custom interferometric inertial sensors.

In this paper we present an initial version of the 6D seismometer detailed in~\cite{MowLowry2019}. The basis behind the design is a softly suspended extended reference mass which is readout in six degrees of freedom (6D). Unlike the inertial sensors discussed above, the approach differs by utilising a simple mechanical design which enables cross couplings. Complexity is moved to the signal processing where the degrees of freedom must be untangled. 

We demonstrate the viability of the device for use in feedback by stabilising a rigid isolated platform in six degrees of freedom. First we discuss the experimental design, and then move through the control scheme, indicating the performance achieved and the shortcomings of the test bed used. 

\section{Optomechanical design}

\begin{figure*}[t]
\centering
\begin{subfigure}{0.48\linewidth}
\centering
\includegraphics[width=\linewidth]{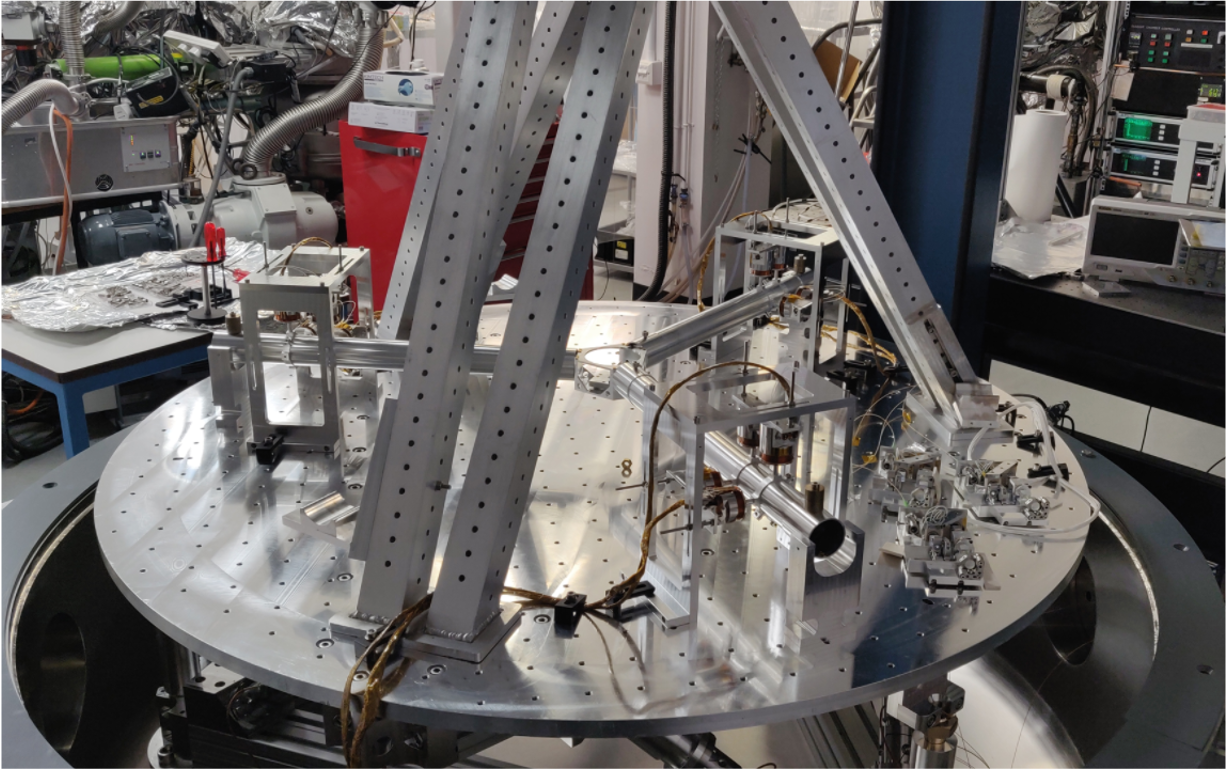}
\caption{Photo of experimental setup. The round active platform is stabilised relative to the seismometer with three arms.}
\label{subfig:setup}
\end{subfigure}
\hfill
\begin{subfigure}{0.48\linewidth}
\centering
\includegraphics[width=\linewidth]{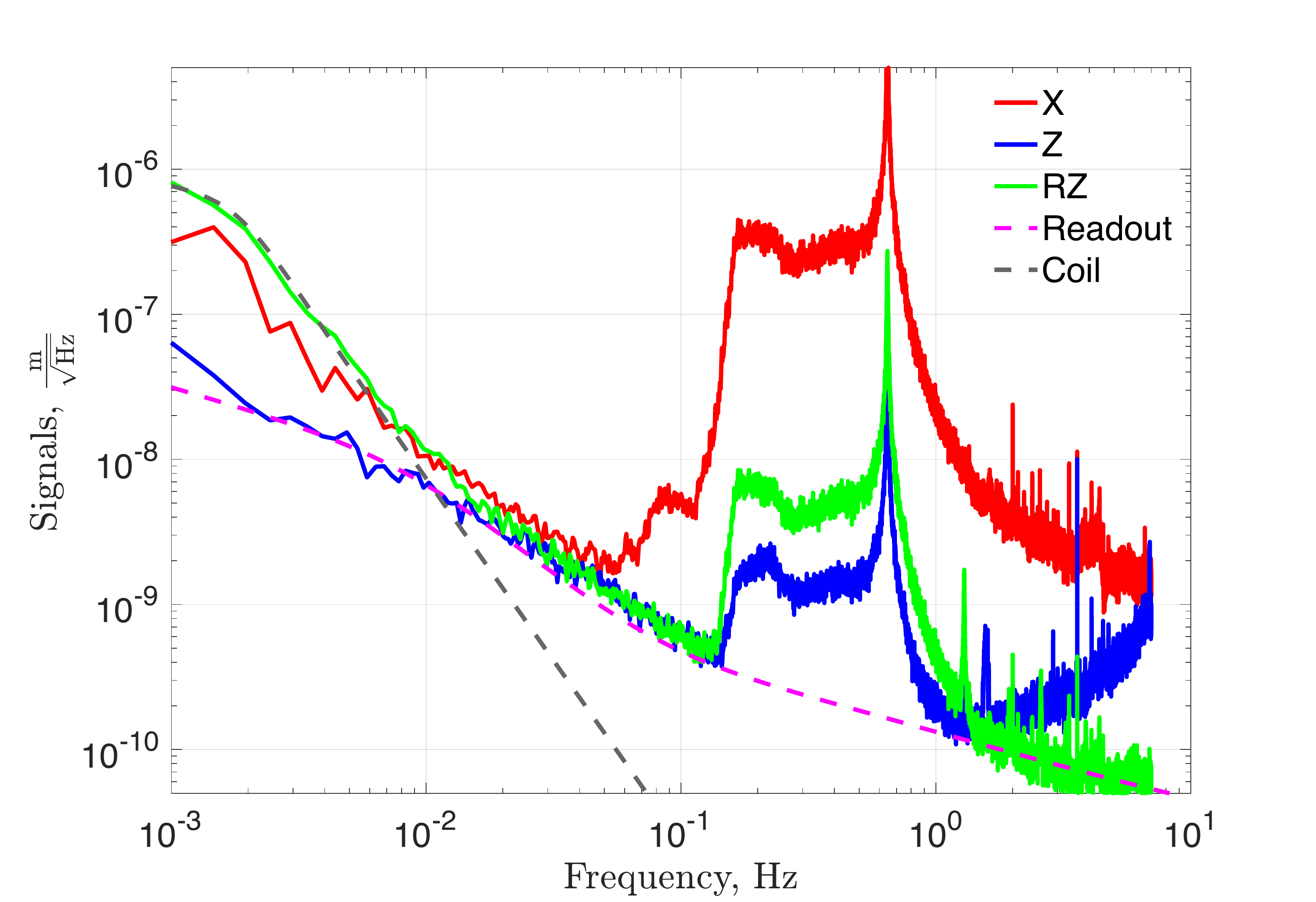}
\caption{Damped signals of the various degrees of freedom to highlight the noise floors due to sensing (Readout) and damping (Coil). Note that the Y and RX degrees of freedom are analogous to X and RY.}
\label{subfig:signals}
\end{subfigure}
\caption{(a) Image of the experimental setup and (b) an example of the measured signals with the corresponding noises.}
\label{fig:readout}
\end{figure*}

The seismometer consists of a single extended reference mass suspended from a fused silica fibre~\cite{Cumming2012,Cumming2020}. Optical shadow sensors known as Birmingham Optical Sensors and Electromagnetic Motors (BOSEMs)~\cite{Strain2012} were employed for the readout scheme, which measured the relative displacement between the proof mass and the platform. The test bed was a rigid stabilisation platform which was actuated using six piezo legs in a hexapod style formation. The experimental set up is shown in Fig.~\ref{subfig:setup} and experimental parameters, highlighting the resonant frequencies of the proof mass, are summarised in Table~\ref{tab:params}.

Ideally the eigenmodes of the mass should be as low as possible to enable inertial sensing to lower frequencies. The stiffest degree of freedom in our setup is the vertical one and the corresponding eigenfrequency of its bounce mode is 10\,Hz. The other two translations degrees of freedom were softer with eigenmodes of 0.62\,Hz. The eigenfrequencies were determined by the fibre length which was constrained by the height of the vacuum chamber.

Resonant frequencies for the tilt modes (RX, RY) were tuned to 100\,mHz and 90\,mHz by compensating the elastic restoring coefficient of the fibre 
with the gravitational anti-spring. The distance between the effective pivot point of the wire and the centre of mass, $d$ enabled tuning of the effective restoring torque as indicated in Eq.~(\ref{eq:resonances})~\cite{Ubhi2022},

\begin{equation}
    \centering
    \omega_{\rm X}^2 \approx \frac{g}{L}, \qquad \omega_{\rm RY}^2 \approx \frac{m g d + k_{el}}{I_y},
    \label{eq:resonances}
\end{equation}
where $m$ mass, $k_{el}$ the elastic restoring coefficient, and $I_y$ is its moment of inertia about the y-axis.

\begin{table}[t]
    \caption{A list of parameters and nominal values}
    \begin{tabular}{ccc}
    \hline
    \hline
    Parameters & Description & Value\\
    \hline
    $m$ & Mass & 1\,kg \\
    \hline
    $R$ & Mass radius & 0.6\,m \\
    \hline
    $L$ & Fibre length & 0.64\,m \\
    \hline
    $r$ & Fibre radius & $100\pm 10\, \mu \rm m$ \\
    \hline
    $f_{\rm X,Y}$ & Translational resonances & 0.62\,Hz\\
    \hline
    $f_{\rm Z}$ & Vertical resonance & 10\,Hz\\
    \hline
    $f_{\rm RX}$ & RX tilt resonance & 100\,mHz \\
    \hline
    $f_{\rm RY}$ & RY tilt resonance & 90\,mHz\\
    \hline
    $f_{\rm RZ}$ & Tilt resonance & 2\,mHz\\
    \hline
    \hline
    \end{tabular}
    \label{tab:params}
\end{table}

The soft angular modes of the system result in large oscillations which ring down over extended periods of time. In particular, the ring down time of the torsion mode (RZ) is several months. In order to maintain the BOSEM sensors within their linear regime, we implemented damping loops on the seismometer's resonant modes using coil-magnet pairs. The damping loops actuated directly on the mass in narrow frequency bands around its resonances and reduce the mass motion down to $\sim \mu$m level.


Fig.~\ref{subfig:signals} shows the damped signals using the BOSEM actuation with no control of the platform. Large translational motion in X leaks into the other degrees of freedom, which can be seen from the presence of the microseism and resonant peak at 0.62\,Hz. Reduction of the X (Y) platform motion diminishes this effect as the platform tracks the motion of the proof mass. Experimental investigations into the BOSEM sensing and actuation noise found that the stiffest mode (Z) was limited by sensor noise below 10\,Hz, and that the digital-to-analog converter noise from our control system dominates the RZ motion below 10\,mHz.

The following sections discuss the control strategy used to stabilise the actuated platform and the issues faced when controlling a multi degree of freedom system. We then look at the achieved performance and suggest improvements for the system.

\section{Control strategy}

\begin{figure*}[t]
\centering
\begin{subfigure}{0.48\linewidth}
\centering
\includegraphics[width=\linewidth]{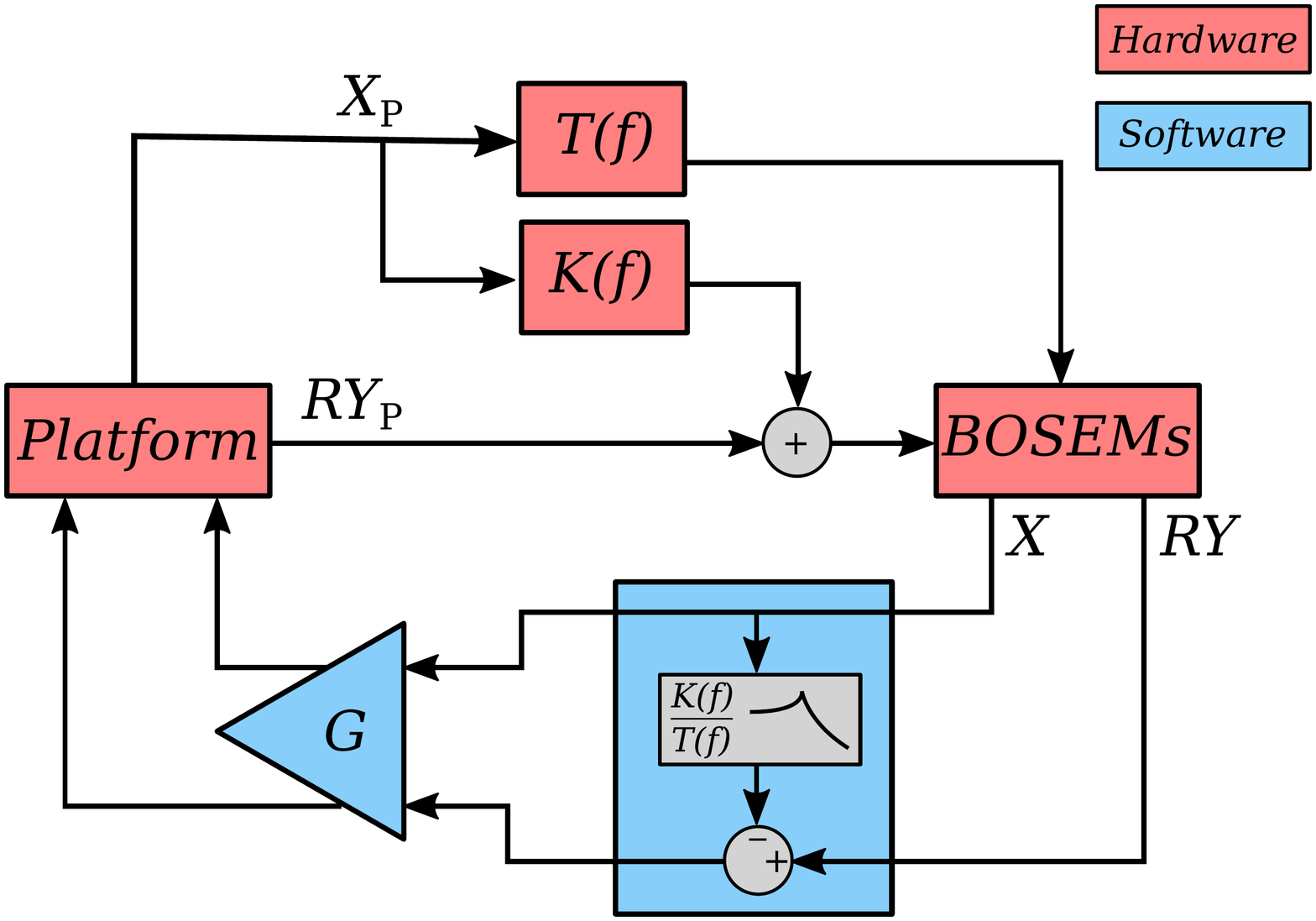}
\caption{Block diagram showing the feedforward scheme to subtract the translational induced tilt motion from the RY measurement.}
\label{subfig:control}
\end{subfigure}
\hfill
\begin{subfigure}{0.48\linewidth}
\centering
\includegraphics[width=\linewidth]{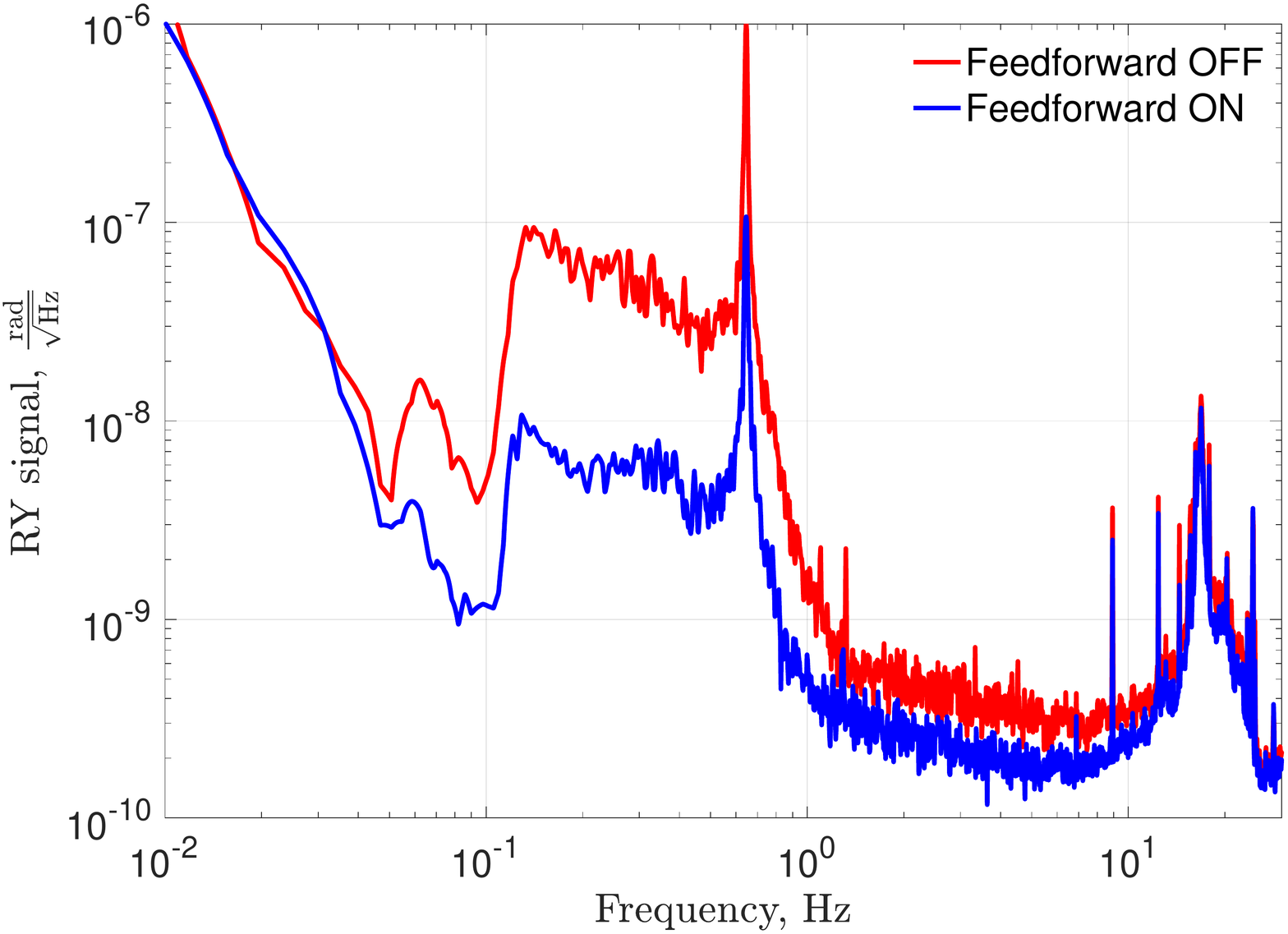}
\caption{Comparision of the RY signal with (blue) and without (red) feedforward of the X signal.}
\label{subfig:feedforward}
\end{subfigure}
\caption{Feedforward scheme for decoupling tilt translation from tilt.}
\label{fig:feedforward}
\end{figure*}

In this section, we discuss the stabilisation technique of the actuated platform relative to the 6D seismometer. First, we present our solution to the control problem. We found that the key element for the successful stabilisation is the feedforward subtraction of the measured longitudinal signals (X and Y) from the tilt signals (RY and RX). Second, we discuss the control problem that is relevant to the class of actuated platforms with cross couplings between different degrees of freedom on the level of $\approx 1$\%.

\subsection{Diagonalisation of the tilt modes}

In the case of a symmetric fibre neck and mass, the circular cross section results in an infinite number of principle axes, resulting in no preferential axes around which the tilt motion occurs. This was initially assumed and an arbitrary direction for the X and Y axes was chosen. 

We discovered a discrepancy between the tilt resonances such that $f_{\rm RX} \neq f_{\rm RY}$. Investigations determined that asymmetry in the fibre neck, where bending occurs, gave rise to two perpendicular principal axes around which tilting occurred. The asymmetry resulted in non-identical elastic restoring constants, $k_{el}$, for RX and RY, where the frequency splitting of the modes was further exacerbated by the tunable gravitational restoring torque, $mgd$. 

Measurement of the degrees of freedom were determined using a sensing matrix, $\mathbf{S}$, which converted the six BOSEM signals, $\vec{B}$ into the six degrees of freedom, $\vec{X}$, such that $\vec{X} = \mathbf{S}\vec{B}$. The preferential axes for tilt caused coupling of the RX eigenmode into the sensed RY motion (and RY to RX). Analysis of the individual BOSEM signals allowed us to determine the angular misalignment of our original axes compared to the principal axes due to the fibre asymmetry. A rotation matrix, $\mathbf{R}$ was implemented to align the sensing with the eigenmodes of the principal axes, $\vec{X}_{eig}$, such that,

\begin{equation}
    \centering
    \vec{X}_{eig} = \mathbf{R}\vec{X} = \mathbf{R}\mathbf{S}\vec{B}.
    \label{eq:rotation}
\end{equation}

Similar to the sensing matrix, the platform actuation was set to align its principle rotation axis with the 6D seismometer.

\subsection{Horizontal-to-tilt decoupling}

\begin{figure*}
    \centering
    \begin{subfigure}[b]{0.475\textwidth}
        \centering
        \includegraphics[width=\textwidth]{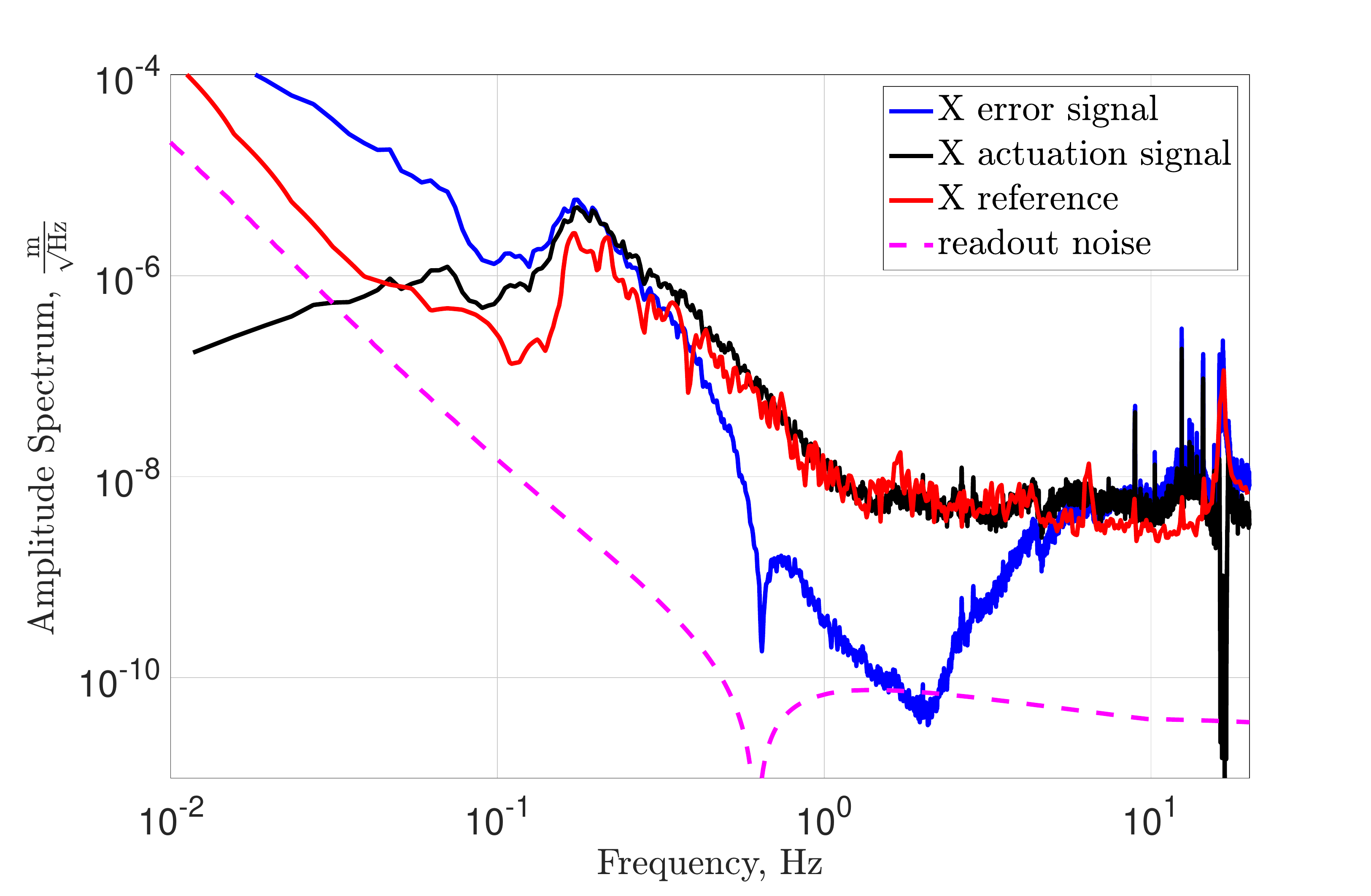}
        \caption{X degree of freedom.}%
        \label{fig:Xstab}
    \end{subfigure}
    \hfill
    \begin{subfigure}[b]{0.475\textwidth}  
        \centering 
        \includegraphics[width=\textwidth]{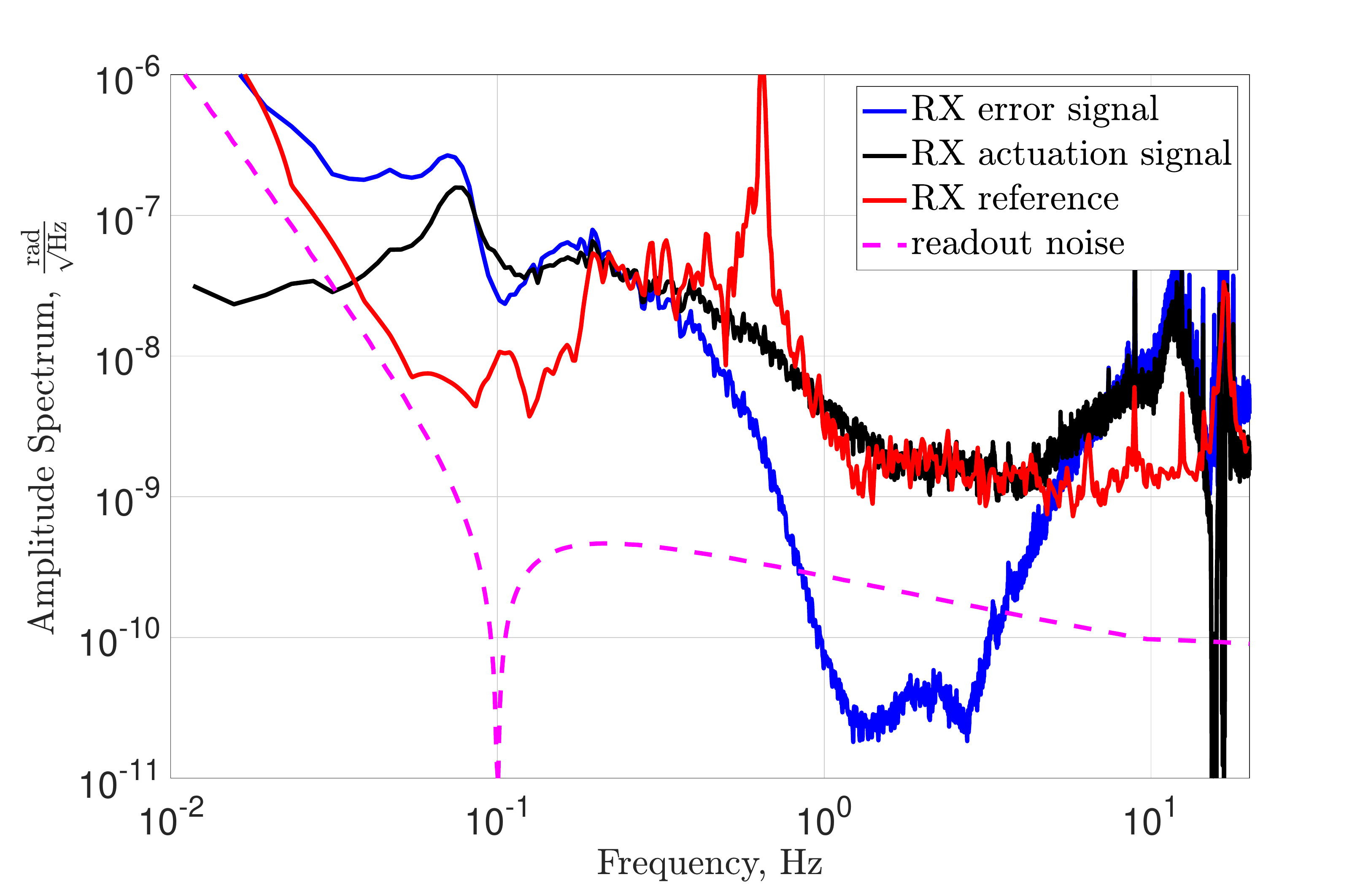}
        \caption{RX degree of freedom.}%
        \label{fig:RXstab}
    \end{subfigure}
    \vskip\baselineskip
    \begin{subfigure}[b]{0.475\textwidth}   
        \centering 
        \includegraphics[width=\textwidth]{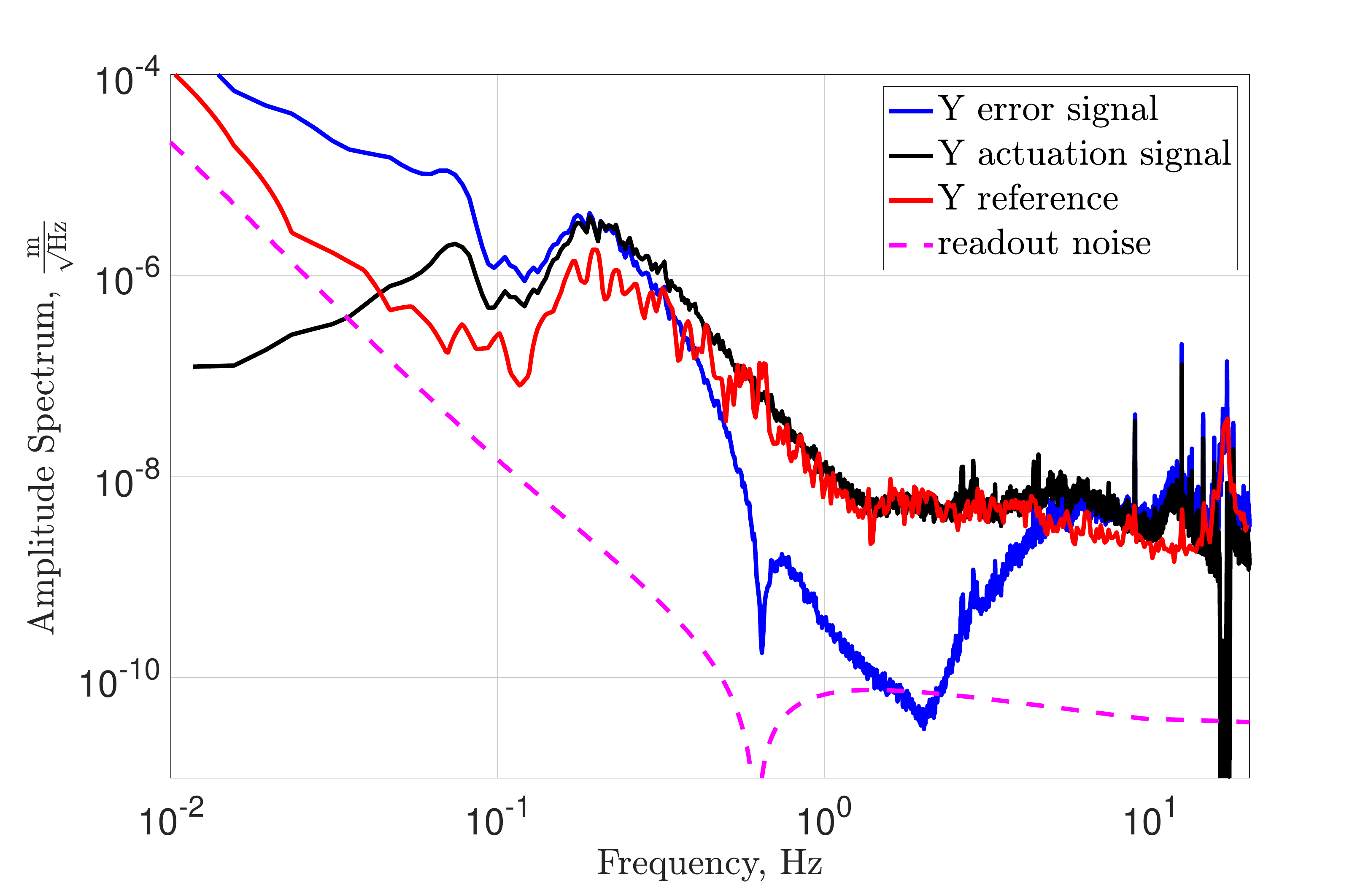}
        \caption{Y degree of freedom.}%
        \label{fig:Ystab}
    \end{subfigure}
    \hfill
    \begin{subfigure}[b]{0.475\textwidth}   
        \centering 
        \includegraphics[width=\textwidth]{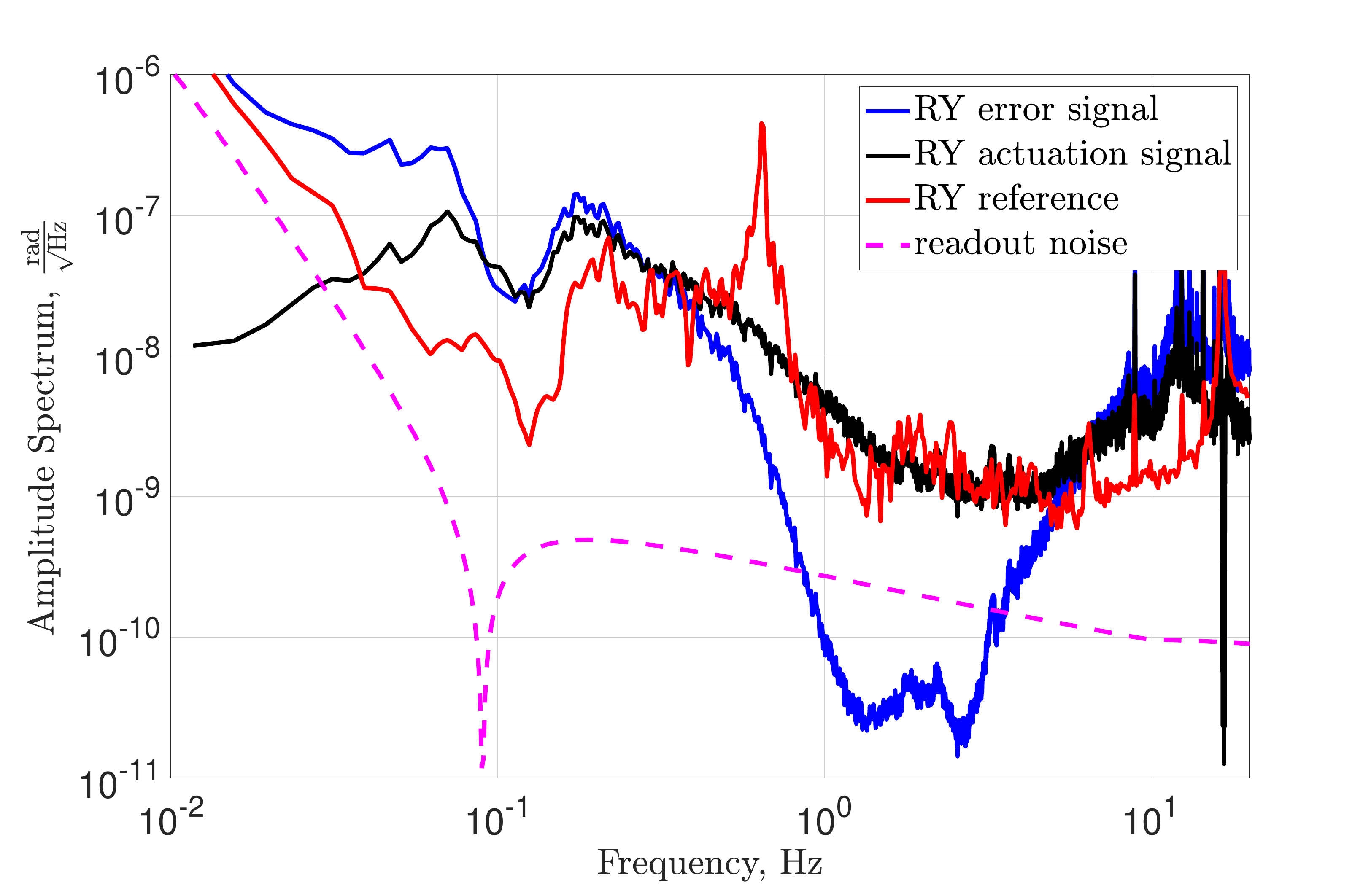}
        \caption{RY degree of freedom.}%
        \label{fig:RYstab}
    \end{subfigure}
    \vskip\baselineskip
    \begin{subfigure}[b]{0.475\textwidth}   
        \centering 
        \includegraphics[width=\textwidth]{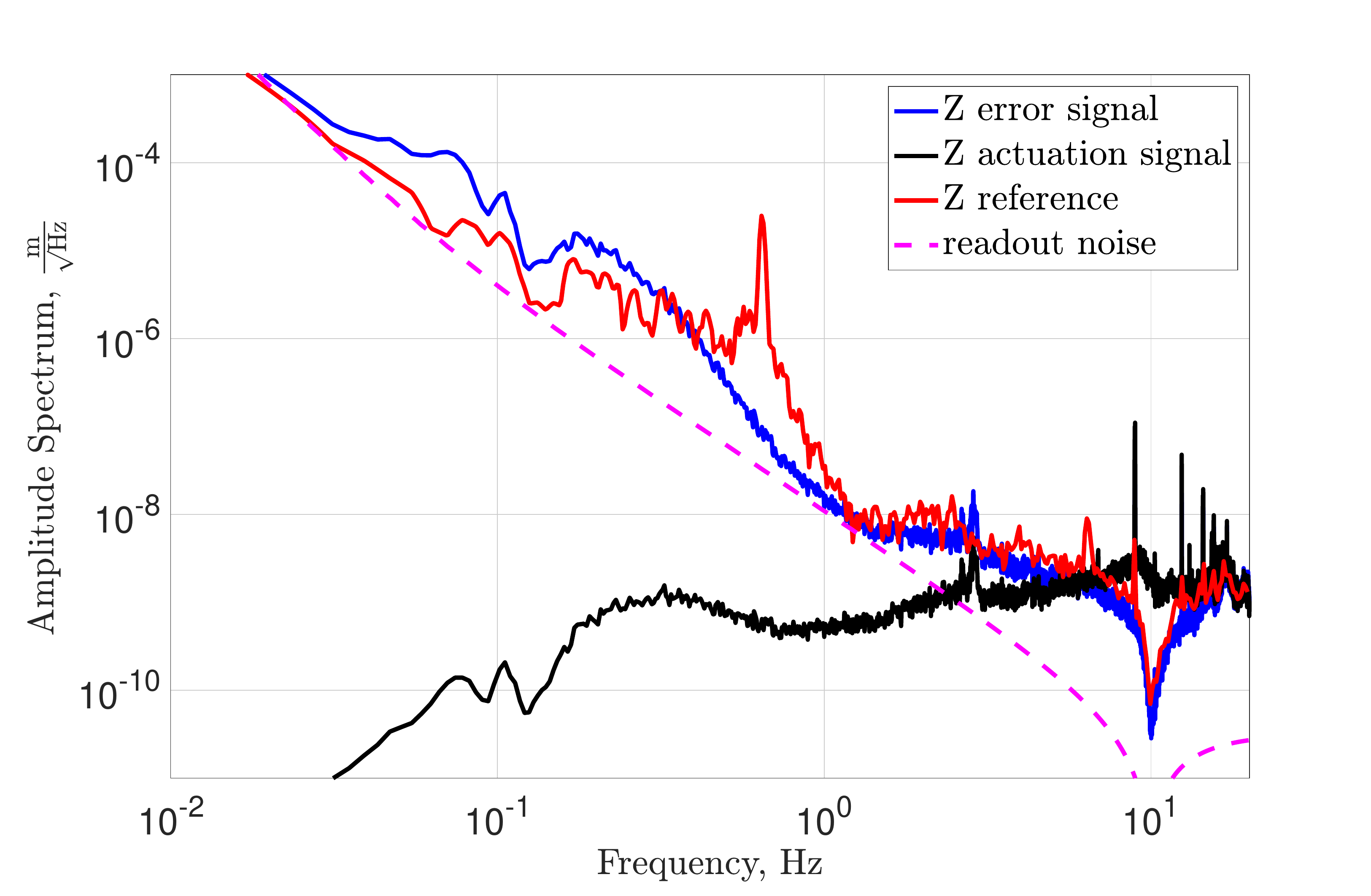}
        \caption{Z degree of freedom.}%
        \label{fig:Zstab}
    \end{subfigure}
    \hfill
    \begin{subfigure}[b]{0.475\textwidth}   
        \centering 
        \includegraphics[width=\textwidth]{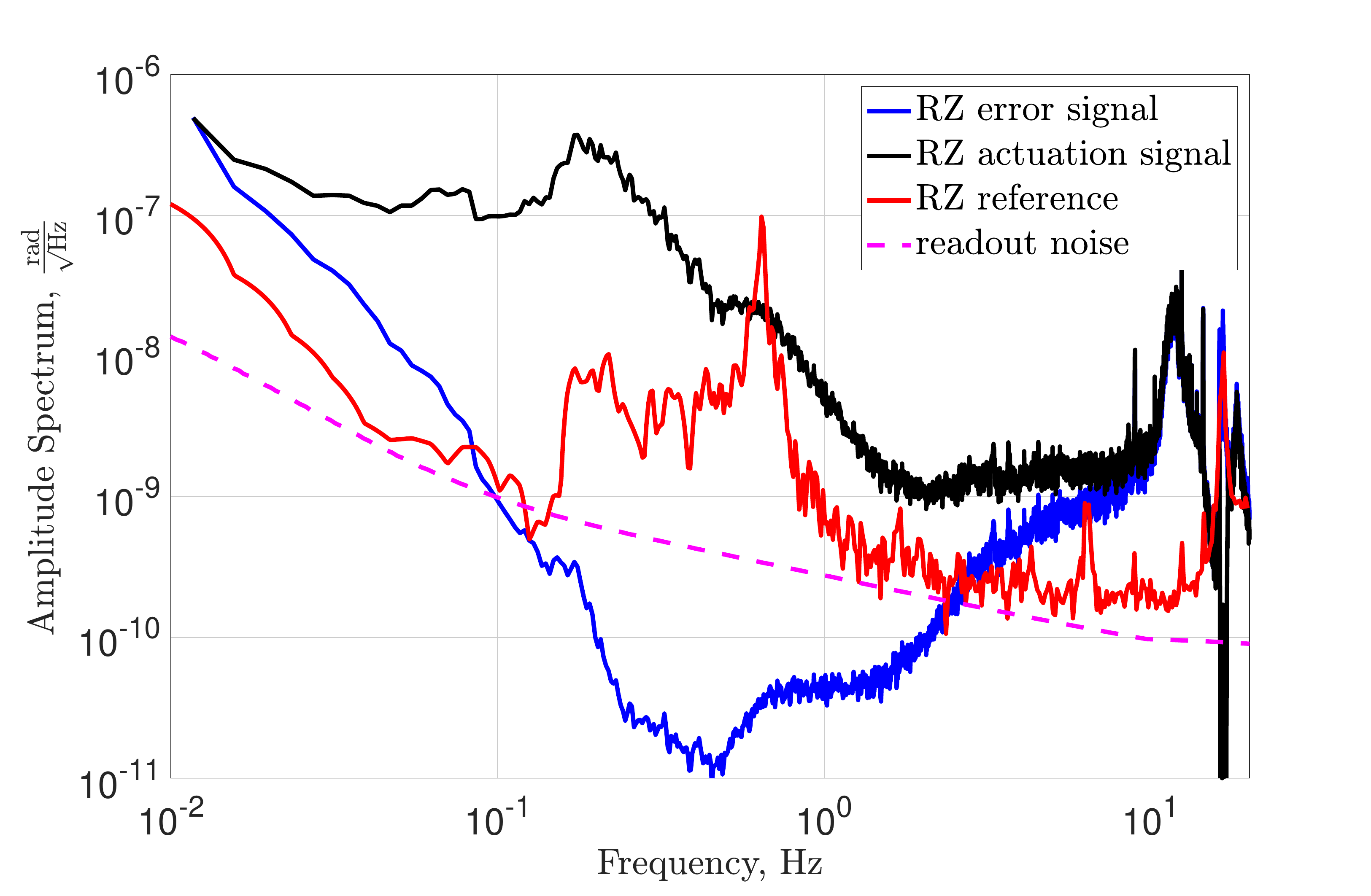}
        \caption{RZ degree of freedom.}%
        \label{fig:RZstab}
    \end{subfigure}
    \caption{Performance of the platform stabilisation using the 6D seismometer for simultaneous control of all six degrees of freedom. } 
    \label{fig:ISIstab}
\end{figure*}


The platform causes movement of the suspension frame and the test mass which is shown in Fig.~\ref{subfig:setup}. However, the test mass is considered to be inertial above the pendulum resonant frequencies. The coupling of platform motion, $X_{\rm P}$ and $RY_{\rm P}$, to the sensor outputs, $X$ and $RY$, can be written as
\begin{equation}\label{eq:isi_tm}
\begin{split}
	& X = T(f) X_{\rm P} + L \times RY_{\rm P} \hspace{1cm}, \\
	& RY = K(f) X_{\rm P} + RY_{\rm P},
\end{split}
\end{equation}
where $L$ is the fibre length. X and RY as well as Y and RX are intrinsically coupled by the pendulum. Transfer functions $T(f)$ and $K(f)$ are determined by the pendulum and pitch resonances and are discussed in details in~\cite{Ubhi2022}.

According to Eq.~(\ref{eq:isi_tm}), the coupling of X and RY and, similarly, Y and RX degrees of freedom is frequency dependent. Therefore, we implement a filter to diagonalise the degrees of freedom as shown in Fig.~\ref{subfig:control}. We found that the control system requires the subtraction of X (Y) from RY (RX), hence a 2x2 diagonalisation is necessary for stability.

We determined the feedforward filter by solving Eq.~(\ref{eq:isi_tm}) relative to $X_{\rm P}$ and $RY_{\rm P}$. Since the solutions are given by the equations
\begin{equation}\label{eq:readout}
\begin{split}
	& X_{\rm P} = \frac{1}{T-L K}\left(X - L \times RY \right), \\
	& RY_{\rm P} = \frac{T}{T-L K}\left(RY - \frac{K}{T}X \right) ,
\end{split}
\end{equation}
the feedforward filter should be given by the equation
\begin{equation}
    \frac{K}{T} = \frac{\omega_{\rm RY}^2}{-\omega^2 + \frac{i \omega \omega_{\rm RY}}{Q_{\rm RY}} + \omega_{\rm RY}^2}\frac{1}{L} \approx -\frac{\omega_{\rm RY}^2}{\omega^2 L}
\end{equation}
at $\omega \gg \omega_{\rm RY}$.
However, during our experimental studies we found that $\sim \omega^{-2}$ dependence is only valid up to $\omega \approx 10 \omega_{\rm RY}$. At higher frequencies, the transfer function flattens due to the direct coupling of horizontal motion to our vertical sensors dedicated for RX and RY. Therefore, we fitted the feedforward filter to the transfer function $K/T + \alpha$, where $\alpha$ is a small number on the order of $10^{-2}$. The result of the feedforward cancellation is shown in Fig.~\ref{subfig:feedforward}.

\subsection{Platform stabilisation}
\label{sec:controlproblems}

Application of the feedforward scheme discussed above enabled successful stabilisation of the platform with 6 single-input-single-output loops. The upper unity gain frequency was constrained to 10\,Hz due to the forest of mechanical resonances of the vacuum chamber and its supporting structure above 14\,Hz. The resonances modify the actuation path of the feedback control scheme, and due to the large number of modes, it was implausible to digitally remove the resonances from all degrees of freedom. The bandwidths achieved for the angular modes were 70\,mHz-10\,Hz for the tilt modes (RX, RY), and 10\,mHz-10\,Hz for RZ. For the longitudinal degrees of freedom (X, Y) the bandwidth attained was 250\,mHz-10\,Hz, where the lower unity gain frequency was limited by the cross-couplings of the platform actuation between the X and Y degrees of freedom.

Above 1\,Hz, the cross-coupling between X and Y degrees of freedom is caused by the imperfect actuation diagonalisation matrix and is on the order of 1\%. However, the coupling grows significantly towards lower frequencies making the response in X and Y to the excitation in X equal at 40\,mHz. 
The large cross-coupling is caused by the tilt-to-horizontal coupling and imperfections of the actuation system: excitation in X also drives RX, resulting in the unpleasant $g/\omega^2$ tilt coupling into the Y degree of freedom. As a consequence, the open loop transfer function of the X degree of freedom is altered when control of Y is simultaneously enganged according to the equation
\begin{equation}
    H_{\rm mod} = H + \frac{\beta_x \beta_y G^2}{1-H}.
\end{equation}
Here, $H = H_x = H_y$ is the open loop transfer function when stabilisation of only one degree of freedom (X or Y) is active, $G$ is the servo gain as shown in Fig.~\ref{subfig:control}. The additional factor is proportional to the cross-coupling of the X degree of freedom to Y, $\beta_x$, and to the similar coefficient from Y to X, $\beta_y$. The additional factor increases the magnitude of the open loop transfer function and makes the closed loop behaviour unstable if the lower unity gain frequency of the feedback loop is below 90\,mHz for $|\beta_y| = |\beta_x| = 10^{-2}$.

We could reduce the actuation imperfections $\beta_x$ and $\beta_y$ down to 0.3\% by gain matching the piezo actuators. However, the hysteresis of the actuators causes time-dependent changes to the gains of the piezos depending on the control system. Since the actuation system is non-linear, we can not reduce the cross-coupling coefficients $\beta_x$ and $\beta_y$ to the levels below 1\% consistently. As a result, we have reduced the control bandwidth in the X and Y degrees of freedom to avoid the instabilities caused by the actuation cross-couplings. However, we expect that the problem of non-linear cross-coupling between X and Y degrees of freedom is not present in the suspended active platforms utilised in LIGO~\cite{Matichard_2015}.

\subsection{Vibration isolation}
\label{sec:performance}

Non-linearity of the actuation path reduced the desired bandwidth of the feedback control system. However, high gain stabilisation of all six degrees of freedom was achieved once correct implementation of the feedforward scheme between X, RY and Y, RX was performed. For the 5 softer degrees of freedom this resulted in two orders of magnitude suppression around 1\,Hz as shown in Fig.~\ref{fig:ISIstab}. 

Vertical suppression was limited due to the stiff resonant frequency, reducing the bandwidth over which stabilisation occurred. Below 1\,Hz the actuation in Z was negligible due to non-inertial sensing which would result in sensor noise injection. Reduction of the resonant frequency can be achieved by suspending the system from a soft blade spring to reduce the bounce mode, or by increasing the tension on the fibre. The Glasgow group are currently developing higher stress fibres for use in third generation detectors~\cite{Cumming2022}.

The majority of the sensed low frequency motion came from the translational modes, X and Y, and were dominated by the microseismic motion between $0.2\,\rm Hz$ and the 0.62\,Hz resonant peaks. The large motion leaked into the other degrees of freedom and can be seen by the red reference traces (no stabilisation) in Fig.~\ref{fig:ISIstab} due to the imperfections of the sensing scheme. Implementation of the feedforward scheme described in Sec.~\ref{sec:controlproblems} suppressed the coupling into the tilt modes by an order of magnitude in the frequency band from 0.1 to 1\,Hz (Fig.~\ref{fig:feedforward}). 

The error signals in Fig.~\ref{fig:ISIstab} show the achievable isolation for the current system, however, this is limited by the BOSEM readout noise highlighted by the magenta traces. Further broadband suppression down to the readout noise level was constrained by the limited bandwidth for all degrees of freedom. Improved readout noise would enable the isolation to be solely limited by the bandwidth and the issues discussed in Sec.~\ref{sec:controlproblems}.

\section{Conclusions}

We have demonstrated the viability of stabilising a six axis platform using a 6D seismometer. The system was operated in high gain with a maximised bandwidth, providing simultaneous control of all six degrees of freedom. We were able to achieve isolation of more than an order of magnitude at 1\,Hz for 5 of 6 degrees of freedom. We found the two key principles of the successful control strategy: sensing diagonalisation of the tilt modes and decoupling of the horizontal-to-tilt motion. The control techniques are a necessity to diagonalise the degrees of freedom involved in feedback control and to make the overall control system stable.

The system can be further improved in three directions. First, the sensing noise of optical shadow sensors can be improved by two orders of magnitude using interferometric inertial sensors~\cite{DFM, Gerberding_2017, Smetana2022compact}. Interferometric sensing has been employed to the system and is currently being optimised to reduce the readout noise.

Second, the system is susceptible to drift motion for the angular degrees of freedom due to thermal gradients, stress relaxations in the fibre and in the metal proof mass. We have acquired a fused silica proof mass (discussed in~\cite{Ubhi2022}) which has the potential to reduce the drift motion of the suspended mass due to its low thermal expansion coefficient and lack of plastic deformations. Themal shielding is also being installed to further isolate the proof mass.

Finally, the actuation of the platform can be improved by suspending it and using coil-magnets actuators similar to the LIGO platforms~\cite{Matichard_2015}. As an intermediate step we may introduce viton sheets to provide passive isolation to damp the high frequency resonant modes of our chamber. This will allow the upper unity gain frequency of the control loops to be increased improving the achievable isolation.


\section*{Acknowledgements}
We thank Rich Mittleman for his valuable internal review and also members of the LIGO SWG groups for useful discussions.
The authors acknowledge the support of the Institute for Gravitational Wave Astronomy at the University of Birmingham, STFC 2018 Equipment Call ST/S002154/1, STFC 'Astrophysics at the University of Birmingham' grant ST/S000305/1, STFC QTFP "Quantum-enhanced interferometry for new physics" grant ST/T006609/1. A.S.U is supported by STFC studentships 2117289 and 2116965.
A.M contributed in the design of the coil magnet actuation scheme for damping of the test mass. A.M, J.V.D, and C.M.L are funded by the European Research Council (ERC) under the European Union's Horizon 2020 research and innovation programme (grant agreement No. 865816).

\section*{Bibliography}
\bibliography{main.bib}

\end{document}